# Strongly Interacting Higgs Sector Without Technicolor

Chuan Liu [a*] with Karl Jansen [b] and Julius Kuti[a]

[a]Department of Physics 0319, University of California at San Diego, 9500 Gilman Drive, La Jolla, CA 92093-0319, USA

[b] DESY -T-, Notkestrasse 85, 22607 Hamburg, Germany

Simulation results are presented on Higgs mass calculations in the spontaneously broken phase of the Higgs sector in the minimal Standard Model with a higher derivative regulator. A heavy Higgs particle is found in the TeV mass range in the presence of a complex conjugate ghost pair at higher energies. The ghost pair evades easy experimental detection. As a finite and unitary theory in the continuum, this model serves as an explicit and simple example of a strongly interacting Higgs sector without technicolor.

## 1. Introduction

Following the systematic method that was developed earlier [1], we report our simulation results on a heavy Higgs particle in the Higgs sector of the minimal Standard Model with a higher derivative regulator. The quantization procedure in Minkowski space-time with indefinite metric Hilbert space for the $O(4)$ symmetric higher derivative Higgs Lagrangian,

$$\begin{aligned}\mathcal{L} &= -\frac{1}{2}\,\Phi_\alpha\Box\Phi_\alpha - \frac{1}{2M^4}\,\Phi_\alpha\Box^3\Phi_\alpha \\ &- \frac{1}{2}\,m_0^2\,\Phi_\alpha\Phi_\alpha - \lambda_0\,(\Phi_\alpha\Phi_\alpha)^2 \;, \qquad (1)\end{aligned}$$

leads to the euclidean partition function [1,9] $\mathcal{Z} = \int\mathcal{D}\Phi\exp\left(-\int d^4x\mathcal{L}_E\right)$, where $\mathcal{L}_E$ is obtained from Eq. (1) by replacing the Minkowski operator $\Box$ with the euclidean Laplacian.

The general particle content of the model can be exhibited by diagonalizing the free Hamiltonian in terms of creation and annihilation operators. The interaction part of the Hamiltonian can be expressed in terms of creation and annihilation operators as well [9].

*Speaker at the Conference, current address: DESY -T-, Notkestrasse 85, 22607 Hamburg, Germany. This work was supported by the DOE under Grant DE-FG03-90ER40546.

## 2. Unitarity

It has been pointed out before that field theories with a complex ghost pair will have a unitary $S$-matrix in the scattering of ordinary particles [8]. The unitarity of the $S$-matrix has been demonstrated within the framework of the $1/N$ expansion in the broken phase of our higher derivative model [10]. Starting from the Lippmann-Schwinger equation, one can show that the elastic scattering amplitude of Goldstone particles remains unitary at all energies and the effects of the ghost pair are hidden in experimentally measurable quantities. In Fig.1, a plot of the scattering cross section and the corresponding phase shift in the isospin $I = 0$ and angular momentum $J = 0$ channel of elastic Goldstone scattering is shown. The peak in the cross section is the standard Higgs resonance in this channel. Surprisingly, as the center of mass energy is increased beyond the energy threshold of the ghost pair, the cross section remains a very smooth function and exhibits no visible signal of the ghost pair. This can be understood from the fact that the complex ghost mass in our theory has a very large imaginary part. If the complex mass of the ghost state were tuned very close to the real axis, one would expect to see a more visible effect on the cross section. Since the $S$-matrix remains unitary, the Higgs model with a higher derivative regulator defines a finite and unitary theory in the continuum whose intrinsic cut-off scale is not revealed by some wrongly assumed pathological properties of the scattering amplitudes.

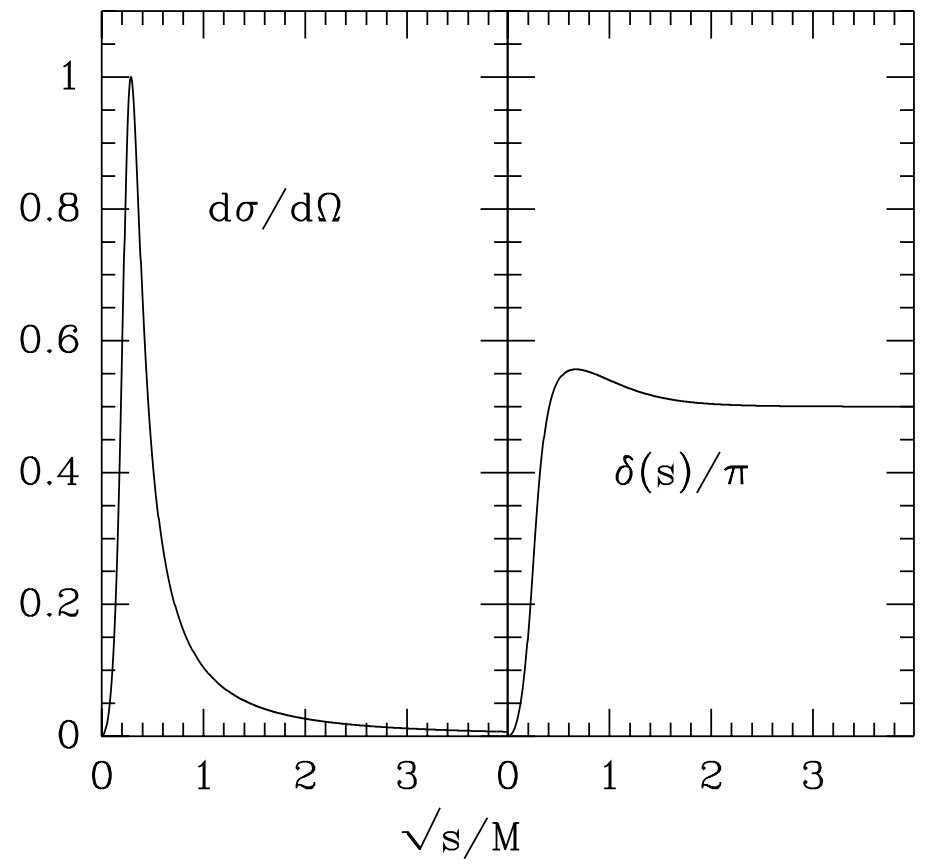

Figure 1. The elastic scattering cross section of two Goldstone particles is shown in the large $N$ approximation as a function of the center of mass energy in the isospin $I = 0$ channel . The peak corresponds to the Higgs resonance. The scattering phase shift of the $I = 0, J = 0$ channel is also shown. Note that the ghost effects are hidden and the scattering is unitary.

## 3. Lattice Action and Phase Diagram

For non-perturbative computer simulations of the higher derivative Lagrangian, we introduce a hypercubic lattice structure. The lattice spacing $a$ defines a new short distance scale with the associated lattice momentum cut-off at $\Lambda = \pi/a$. We will have to work towards the large $\Lambda/M$ limit in order to eliminate finite lattice effects from the already regulated and finite theory. The euclidean lattice action that we studied has the form

$$S_E = \sum_x [-\kappa\ \phi_\alpha(\Box + \frac{\Box^3}{M^4})\phi_\alpha + (1-8\kappa)\phi_\alpha\phi_\alpha + \lambda\ (\phi_\alpha\phi_\alpha - 1)^2]\ , \quad (2)$$

where $\Box$ is the lattice Laplace operator, with the lattice spacing $a$ set to one for convenience.

The lattice model exhibits two phases, as expected. The symmetric phase with full $O(4)$ symmetry is separated by a second order phase transition line from the broken phase which has a residual $O(3)$ symmetry for every fixed value of $\lambda$ in the $(\kappa, M)$ plane.

We developed a Hybrid Monte Carlo algorithm with Fourier acceleration which solved the problem of critical slowing down very effectively for our model [9]. Since the updates are performed in momentum space, we can use improved lattice propagators without significant cost in computer time. This technique greatly extends the parameter range that we can study. Our large $N$ calculations indicate that lattice cut-off effects in the renormalized coupling constant of the improved lattice action are negligible even at $M = 2a^{-1}$.

## 4. Results and Discussion

Most of our computer simulations were performed on lattices which have cylinder geometry in the size range $16^3 \times 40$ to $20^3 \times 40$. On each lattice with a given parameter set, 20,000 to 60,000 Hybrid Monte Carlo trajectories were accumulated. Typical autocorrelation times in our simulations are of the order of 10 trajectories.

A simultaneous fit to $m_H$ and $v$ was performed on the data, as described in [9]. The results are shown in Table 1 where points marked with $*$ correspond to a cylinder size of $20^3 \times 40$. All other points correspond to size $16^3 \times 40$. The last two points were obtained with the improved lattice action while the simple lattice action was applied at the other points.

Table 1

| $\kappa$ | $M$ | $\lambda$ | $v_r$ | $m_H$ | $R$ |
|---|---|---|---|---|---|
| 0.056 | 0.8 | 0.4 | 0.057(2) | 0.40(2) | 7.0(4) |
| 0.056 | 0.8 | 0.4 | 0.045(1) | 0.33(2) | 7.3(5)* |
| 0.105 | 0.8 | 0.1 | 0.065(2) | 0.31(2) | 4.8(4) |
| 0.115 | 0.8 | 0.05 | 0.082(1) | 0.24(1) | 2.9(1) |
| 0.081 | 1.0 | 0.3 | 0.093(1) | 0.42(2) | 4.5(2) |
| 0.081 | 1.0 | 0.3 | 0.088(1) | 0.38(2) | 4.3(3)* |
| 0.088 | 2.0 | 1.0 | 0.058(1) | 0.351(5) | 6.1(1) |
| 0.088 | 2.0 | 1.0 | 0.045(1) | 0.29(1) | 6.4(3)* |

The radial Higgs excitation in our simulations corresponds to a stationary state with real energy. Avoided level crossing with the lowest Goldstone pair of zero total momentum would only occur in larger spatial volumes. However, both the large $N$ calculation and the practical simulation data seem to indicate that the ratio $m_H/v_r$ is not sensitive to the size of the volume in the range of our simulations. A more sophisticated study of the Higgs resonance in the finite volume utilizing Lüscher's prescription is feasible in our model. However, preliminary runs show that one needs significant computer resources to get reasonably accurate results for the scattering phase shift.

The ghost mass parameter quoted in the table is the bare value. It is very hard to determine the precise ghost pole location from the Higgs correlation function due to its mixing with the Higgs channel. However, fits to the momentum space propagator and analytic calculations in the $1/N$ expansion indicate that renormalization effects in the ghost mass are not significant.

The lattice effects of our simulation points were also analyzed using the method described in [11]. We find that all our simulation points have less than one percent Euclidean violation in the propagator, and they are particularly small for the improved action. Therefore, the results should represent, to a good approximation, the physics described by the continuum higher derivative Higgs model.

It is seen that the $m_H/v_r$ ratio in our model ranges from 3 to 7, and even larger, to be compared with the earlier ratio of 3, or less [2–6]. The Higgs mass range $m_H \geq 1$ TeV, with a ghost location in the multi-TeV region, implies the existence of a strongly interacting Higgs sector in the model.

When viewed as a regulator of the conventional theory, the effects of the complex ghost parameter can be determined in the weak interaction regime. The scattering amplitude can be calculated in the presence of the ghost pair and compared with the corresponding result in conventional perturbation theory. A comparison can also be made within the large $N$ approximation. The large $N$ results show that, for a $m_H/v_r$ ratio of 3, the ratio $M/m_H$ is of the order of 30 at infinite bare coupling. For such a large ghost mass parameter, the $M$ dependence in the scattering amplitude is practically invisible. Therefore, we have to conclude that the Higgs mass bound at around 700 GeV in the old calculations was imposed by the underlying lattice structure. Heavier Higgs mass values can be obtained with the higher derivative regulator without any loss of euclidean invariance, and the theory can be driven into the strongly interacting regime.

This strongly interacting scenario in our model is also confirmed by simulations in the symmetric phase. We have studied two points in the symmetric phase and measured the renormalized coupling constant. For the point with a Higgs mass of $m_H = 0.434$, at $M = 0.8$, we get $\lambda_R = 4.1(7)$. For the point with $m_H = 0.352$, at $M = 2.0$, we get $\lambda_R = 2.2(2)$. These numbers are to be compared with $\lambda_R \approx 1$ for the conventional model at comparable Higgs mass values [3,7].

The Higher derivative Higgs model is a somewhat arbitrary extension of the Higgs sector in the minimal standard model by introducing a new parameter $M$ which represents the threshold of new physics. Since the ghost pair evades easy experimental detection without violating unitarity, or Lorentz invariance, it serves as a toy model of the strongly interacting Higgs sector without technicolor, a scenario which was excluded in previous lattice studies.